\documentclass{mn2e}
\usepackage[dvips]{graphics}

\newif\ifAMStwofonts

\title{Silicate-break galaxies: an efficient selection 
method of distant ultraluminous infrared galaxies}

\author[T. Takagi, C. Pearson]
 {T. Takagi$^{1}$\thanks{E-mail: t.takagi@kent.ac.uk}, 
 C. P. Pearson$^{2}$\\ 
$^1$ 
Centre for Astrophysics and Planetary Science, University of Kent,
Canterbury, Kent, CT2 7NR \\ 
$^2$ 
Japan Aerospace Exploration Agency, Institute of Space and Astronautical Science, 
3-1-1 Yoshinodai, Sagamihara, Kanagawa, Japan
}

\date{}

\begin{document}

\maketitle

\begin{abstract}
We present a photometric selection method for ULIRGs 
in the redshift range of $z=1$ -- 2.  
We utilize the most prominent spectral feature of ULIRGs, 
i.e. the silicate absorption feature at 9.7 $\mu$m and an 
optimized filter system at mid-infrared wavelengths. 
These  
`{\it silicate-break}' galaxies could be selected by using colour 
anomalies owing to the silicate absorption feature around 
9.7(1+$z$) $\mu$m.  Such filter systems are available on the 
{\it Spitzer Space Telescope} 
but we suggest that the most promising
selection criteria would be given with mid-infrared bands of 
{\it ASTRO-F} satellite due to a more comprehensive set of filter 
bands than that of {\it Spitzer}. 
We study the selection method of silicate-break galaxies with 
the SED model of both starbursts and quiescent star-forming galaxies, 
and then verify the method by using the observed spectra of 
nearby galaxies. We would expect that about 
1000 candidates of silicate-break galaxies could be detected 
per square degree in current and future mid-infrared surveys. 
The silicate-break selection criteria will break the degeneracy 
between various galaxy evolution models for extragalactic source counts
and place strong limits on the star formation activity at $z=1$ -- 2. 
Applying our silicate-break technique to initial Spitzer results 
we have tentatively identified the first candidate silicate-break 
galaxy at  $z=1.6$. 
\end{abstract}


\section{Introduction}\label{sec:introduction}

Extreme examples of high-redshift counterparts of nearby 
Ultra-Luminous Infra-Red Galaxies (ULIRGs, ~\cite{sand96}) have been 
found by observations with SCUBA (the Submillimetre Common-User Bolometer 
Array) at 850$\mu$m (e.g. Smail et al.~\shortcite{smail97}, 
Hughes et al.~\shortcite{hugh98}). 
The star formation rates (SFRs) 
of submillimetre (submm) galaxies are estimated to be more than 
$10^3 M_\odot$ yr$^{-1}$ (e.g. Smail et al. ~\shortcite{smail02}, 
Chapman et al. ~\shortcite{chapman03} ), i.e. significantly 
higher than that of nearby ULIRGs. 
The observed flux at submm wavelengths is insensitive to the classical flux-redshift relation 
at $z\ga 1$, owing to the so-called negative K-correction.
Nevertheless, the redshift distribution of submm galaxies 
seems to be rather narrow with a median redshift of $\sim$2.4 
(Chapman et al. ~\shortcite{chapman03} ). 
A tentative detection of strong clustering within the submm population 
suggests that submm galaxies may be the progenitors of today's giant 
elliptical galaxies (Blain et al. ~\shortcite{blain04}). 
A further clue to the evolutionary link 
between submm galaxies and ellipticals is presented by 
the Spectral Energy Distribution (SED) 
fitting analysis by Takagi et al. (2004). 

Spheroids contain about 50 -- 70 \% of the stellar mass in 
the local universe (e.g Schechter \& Dressler ~\shortcite{schecter87},
Fukugita, Hogan, \& Peebles ~\shortcite{fukugita98}). 
It is not yet clear whether the formation of such a large mass fraction can 
be explained by submm galaxies alone, including the recently found 
optically-faint radio galaxies at similar redshifts (Chapman et al. 2004), 
which could have the similar SFR with that of submm galaxies. 
The SFR density due to submm galaxies and optically-faint radio galaxies 
has a strong peak at $z=2$ -- 3 (Chapman et al. 2004), 
which corresponds to the time-scale of only $\sim$1 Gyr. 

Recent studies on near-IR selected galaxies in the HDF-N ~\cite{dickinson03}
 and the HDF-S  ~\cite{fontana03} show that 
a significant fraction of stars at the present epoch 
are formed at $z= 1$ -- 2, i.e. at later epochs than 
when most of submm galaxies are found. 
Therefore, ULIRGs at $z=1$ -- 2 
could be a key galaxy population in understanding the formation 
process of spheroidal galaxies as well as submm galaxies at $z \ga 2$. 
Furthermore it is probable that these infrared galaxies are the main 
contributors to the cosmic infrared background 
(c.f. Chary \& Elbaz ~\shortcite{chary01}). 
Such galaxies will be prime targets of new infrared 
satellites, such as {\it Spitzer} ~\cite{werner04} and 
{\it ASTRO-F}~\cite{cpp04a},~\cite{murakami98}.

Various methods for the photometric pre-selection of interesting 
high-$z$ objects, such as Lyman-break galaxies ~\cite{steidel03}, 
massive galaxies at high redshifts 
(van Dokkum et al. 2004; Cimatti et al. 2004), Lyman-$\alpha$ emitters 
at $z \ga 4$ (Cowie \& Hu 1998; Taniguchi et al. 2003; Shimasaku et al. 2003), 
have revolutionized the field of the galaxy evolution at high redshift. 
These techniques are 
widely used to select objects for spectroscopy, and 
for statistical studies to derive luminosity functions and spatial 
correlation functions, which require a large sample at 
a similar redshift.  

Until now, these photometric pre-selection techniques have been developed 
for galaxy populations observed at optical and near-infrared (NIR) 
wavelengths. 
Similar techniques for selecting luminous infrared galaxies at $z=1$ -- 2 
are strongly demanded. Recently, the spectroscopic bump at 
1.6 $\mu$m due to the H$^-$ opacity minimum has been applied to galaxies 
detected at 24$\mu$m with {\it Spitzer} (Le Floc'h et al. 2004; 
Egami et al. 2004). This technique requires coordinated surveys at 
NIR wavelengths targeting the rest-frame 1.6$\mu$m and MIR 
wavelengths to detect significant dust emission, and also unambiguous 
cross-identification between NIR and MIR sources, which is not 
easy for blended pairs of NIR sources 
(e.g. Egami et al. 2004). Therefore, any techniques 
using only the dust emission would be more efficient.

In the spectra of ULIRGs, the most prominent features are 
found at MIR wavelengths, i.e. silicate absorption at 9.7 $\mu$m and 
the PAH features. 
Here, we investigate the possibility 
of selecting luminous infrared galaxies at $z\ga 1$ with {\it Spitzer} 
and {\it ASTRO-F} by focusing on the MIR features. 
We hereafter refer to distant infrared galaxies selected 
by this method as {\bf `silicate-break}' galaxies for simplicity. 
Note that Charmandaris et al. (2004) briefly discuss 
the possibility of estimating redshifts from the MIR features 
by using 
the Infrared Spectrograph (IRS, Houck et al. ~\shortcite{houck04} )
peak-up imagers at 16 and 22 $\mu$m onboard {\it Spitzer}, 
although it is difficult to use these imagers for large area 
blank field surveys, because of the small filed-of-view. 

\begin{figure}
  \resizebox{\hsize}{!}{\includegraphics{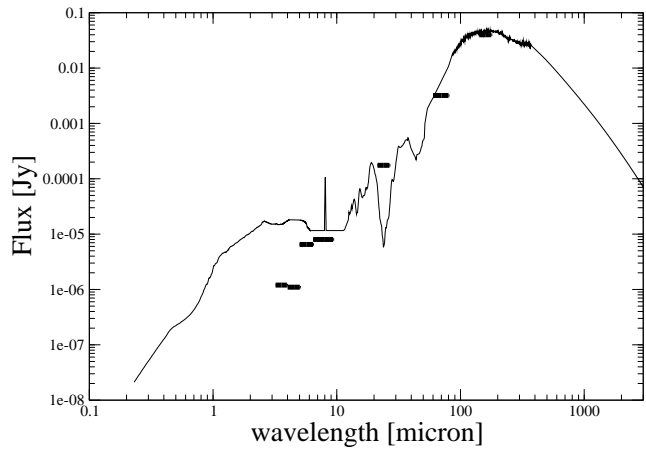}}
 \caption{An example of a potential silicate-break galaxy. 
The SED of Arp220 (Elbaz et al. 2002) at $z=1.5$ is compared with the 
5 $\sigma$ detection limits at {\it Spitzer}/IRAC bands and 
MIPS 24 $\mu$m achieved in the Lockman Hole region 
(Huang et al. 2004; Floc'h et al. 2004). At 70 and 160 $\mu$m, 
we adopt the confusion limit based on the MIPS observations 
(Dole et al. 2004). With these detection limits, galaxies with 
the SED of Arp220 are detectable at all the {\it Spitzer} bands,
except for 24 $\mu$m. 
Note that the luminosity is assumed to be twice of Arp220.
}
 \label{example}
\end{figure}

Silicate-break galaxies could be selected by using the colour 
anomaly owing to the silicate absorption feature 
around a $9.7 (1+z)~ \mu$m band. For example, a galaxy 
with the SED of Arp 220 at $z= 1.5$ would be too faint to be detected 
with {\it Spitzer} in the MIPS 24 $\mu$m band, 
while it would be detectable in the other infrared bands, as 
shown in Figure \ref{example}.
Thus, sources which are detected in all the {\it Spitzer} bands
but 24 $\mu$m are 
candidates for heavily obscured galaxies at $z=1.5$. 
With the sensitivity of {\it Spitzer} and {\it ASTRO-F}, 
most of the silicate-break galaxies would be 
classified as ULIRGs. 

The structure of this paper is as follows. In 
section ~\ref{sec:templates} we describe the SED model used 
to derive the selection criteria for silicate-break galaxies. 
In section ~\ref{sec:selection} we discuss the selection 
criteria for the {\it Spitzer} and {\it ASTRO-F} infrared 
satellite missions and predicted effectiveness of the filter 
combinations in detecting potential silicate-break candidates. 
The potential number of silicate-break galaxies at 
$z =1$ -- 2 are estimated using two infrared evolutionary models 
in section~\ref{sec:numbers}. We give discussion and conclusions in 
section~\ref{sec:conclusions}. Throughout this work we 
assume a flat cosmology of $\Omega_m =0.3$, 
$\Omega_\Lambda =0.7$, and $H_0=75$ km sec$^{-1}$ Mpc$^{-1}$.

\section{SED templates}\label{sec:templates}

In order to derive the colour criteria 
to select silicate-break galaxies in a desired redshift range only, 
we need to ensure that there is negligible contamination from the 
other redshifts by using a wide variety of possible SEDs. 

We adopt the evolutionary SED model of starbursts by Takagi et al. 
(2003a), in which the SED variation is explained 
by the difference in the starburst age and the compactness of the starburst 
region $\Theta$. It is found that the SED model 
reproduces not only the SED of ULIRGs, but also the SED of UV-selected 
starburst galaxies which are usually less luminous. This means that 
the SED model is capable of covering the wide variation of observed 
starburst galaxy SEDs. 

In Takagi et al. (2003a), three types of dust model are adopted, i.e. 
the Milky Way (MW), Large Magellanic Cloud (LMC), and Small Magellanic 
Cloud (SMC) dust models (see Takagi et al. 
2003b for details of the models). The fraction of 
silicate dust grains is assumed to increase from the MW to SMC model, i.e. as 
a function of the metallicity. Therefore, the silicate absorption 
feature is most prominent in the SEDs described by the SMC type dust model. 
Thus, in this work, we focus on the SMC dust model, which is 
found to be suitable for 
most of the nearby ULIRGs modelled by Takagi et al. (2003a). 
We have confirmed that none of the SED models with the MW and LMC type 
dust causes any contamination in the selection criteria discussed below. 

\begin{figure}
\resizebox{\hsize}{!}{\includegraphics{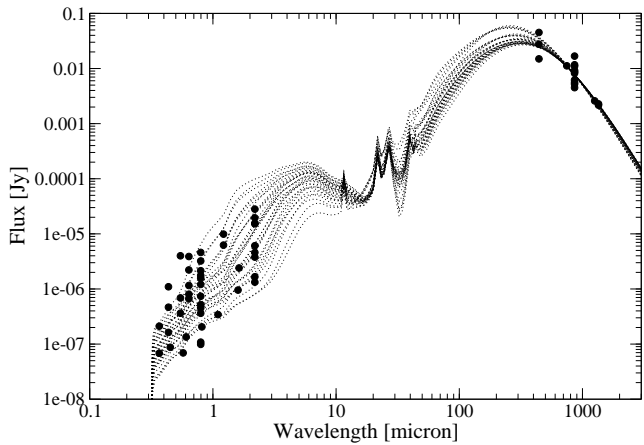}}
\caption{The SED models with $t/t_0 =1$ -- 6 and $\Theta = 0.6$ -- 1.4 
adopted from Takagi et al. (2004). The SMC dust model is adopted. 
The SEDs are normalized with $S_{850} = 8$ mJy at $z=2.5$. 
The observed fluxes of submm galaxies (sold circles) 
are summarized in Takagi et al. (2004). 
}
 \label{SED}
\end{figure}

We constrain the possible parameter space of the SED model, 
 suitable for high-$z$ ULIRGs 
by using the observed 
SED variation of submm galaxies. 
Note that the SED model itself can cover a wide variety 
of starburst SEDs, including that of UV-selected starburst galaxies, 
i.e. non-ULIRGs. 
We first constrain the starburst age at $t/t_0 \ge 1$
where $t_0$ is the evolutionary time-scale of starbursts, 
since the 
probability to select very young
galaxies is low. 
We use the models younger than $t/t_0 = 6$, which are 
enough to reproduce the majority of observed SEDs 
(Takagi et al. 2003a; Takagi et al. 2004). 
We then constrain the compactness $\Theta$ to 0.6 -- 1.4, 
which is enough to cover the observed 
range of SEDs of submm galaxies. 
In Figure \ref{SED}, we show the 
SED models with the starburst age of
$t/t_0 = 1$ -- 6, and $\Theta =0.6$ -- 1.4, together 
with the observed fluxes of submm galaxies at $z= 2$ -- 3. 
The SED models are normalized at $S_{850} = 8$ mJy at $z=2.5$. 
We adopt an initial mass function (IMF) with a power-law index 
of $x=1.10$, which is slightly 
flatter than the Salpeter IMF ($x=1.35$), following Takagi et al. (2004). 
Note that submm galaxies which are very faint 
at optical -- NIR wavelengths,  are not 
included in the sample. Such galaxies could be more heavily obscured 
than the coverage of the SED model, and would therefore have an even deeper 
silicate absorption feature. 

Using the high-$z$ galaxy sample, rather than well studied nearby galaxies, 
to constrain the SED parameter 
space has two distinct advantages; 1) the redshift of the target galaxies
themselves is high ($z>1$), and 2) the effect of any underlying stellar 
population will be minimal in the optical -- NIR in the observed frame. 
Note that the latter advantage is useful in estimating the variation of the
SED parameters from the observed optical -- NIR SEDs.  SED analysis 
of individual submm galaxies is given in Takagi et al. (2004). 

We also need an SED template for quiescent star-forming galaxies to 
estimate any possible contamination of the sample of silicate-break 
galaxies. We adopt the phenomelogical SED template of Dale et al. (2001), 
which reproduces the empirical spectra and infrared colour trends. 
In this model, SEDs are characterized by a power-law index $\alpha$ 
of the distribution of dust mass over a wide range of interstellar 
radiation field strengths. We use the SED models with 
$\alpha \ge 1.25$ which are suitable for the quiescent population.

Both of SED templates by Takagi et al. (2003a) and Dale et al. (2001) 
do not include the contribution of AGN. 
AGN generally produce a featureless continuum which can be 
approximated by a power-law spectrum without PAH emission 
(e.g. Laurent et al. 2000). 
This means that the presence of an AGN will decrease the prominence of the silicate absorption feature. 
Such AGN-dominated galaxies would be difficult to 
be selected as silicate-break galaxies even with high luminosity. 
Therefore, the {\it number} of silicate-break galaxies depends on how 
strong the AGN contribution is in ULIRGs at $z\sim 1.5$. 
However, the AGN contribution is not very important to derive 
the {\it selection criteria} of silicate-break galaxies, since 
AGN-dominated galaxies have almost constant MIR 
colours, owing to the power-law like spectrum. 
This is confirmed below by using observed spectra of AGN in 
section \ref{sec:obssed}. 

All fluxes are calculated using the transmission curves at each 
filter band. The digital form of 
the transmission curves of {\it Spitzer}
bands and {\it ASTRO-F} bands are 
available from the web page of Spitzer Science Center (SSC)
and H. Matsuhara (2004, private communication), respectively, 
except for the {\it Spitzer}/IRS filters at the time of writing. 
For the IRS filters at 16 and 22 $\mu$m, 
we assume box car profiles for the transmission curve.

\section{Selection criteria of silicate-break galaxies}\label{sec:selection}

\subsection{Case for {\it Spitzer}}
\begin{figure}
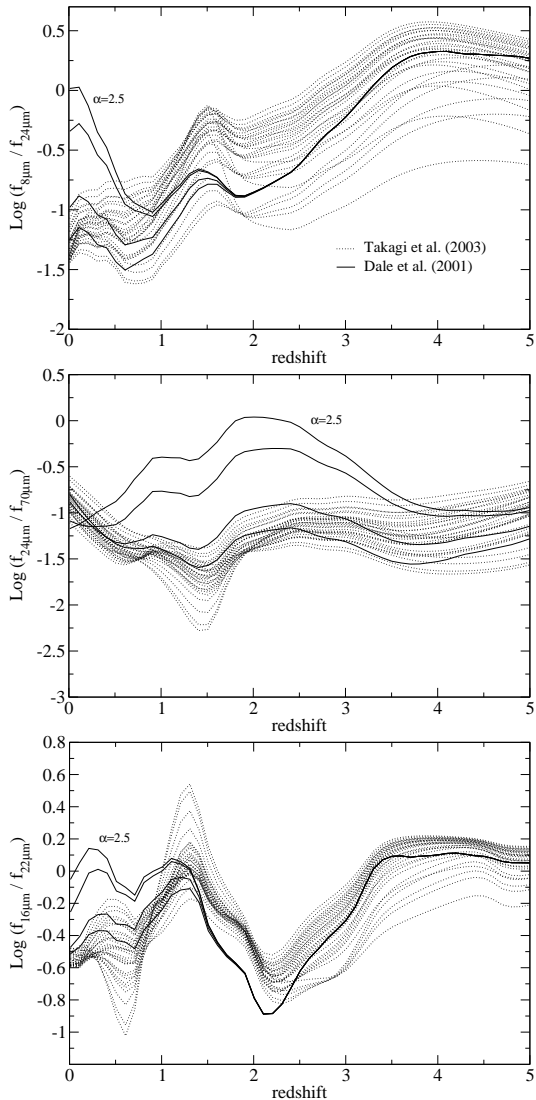

\resizebox{7cm}{!}{\includegraphics{figure/spitzer_8um_24um_smc.eps}}
\resizebox{7cm}{!}{\includegraphics{figure/spitzer_24um_70um_smc.eps}}
\resizebox{7cm}{!}{\includegraphics{figure/spitzer_IRS_smc.eps}}
 \caption{The flux ratios with {\it Spitzer} bands as a function 
of redshift. Dotted lines indicate the SED models by Takagi et al. (2003a), 
shown in Figure \ref{SED}. 
Thin solid lines indicate the SED template by 
Dale et al. with $\alpha = 1.25$, 1.5, 2.0, and 2.5.
The model with $\alpha = 2.5$ (coolest one) is indicated 
explicitly. 
}
 \label{spitzer}
\end{figure}

{\it Spitzer} covers the near to far-infrared range in 7 
bands at 3.6, 4.5, 5.8, 8.0$\mu$m with the IRAC instrument 
~\cite{fazio04} and  
24, 70, 160 $\mu$m with the MIPS 
instrument ~\cite{rieke04}. 
Also, the IRS has peak-up imagers at 16 and 22 $\mu$m, 
although the field-of-view is small. These imagers would be 
useful for targeted observations of objects which are too faint to 
obtain spectroscopic redshifts. 
In Figure \ref{spitzer},   we show the three 
flux ratios, 8-to-24$\mu$m, 24 to 70$\mu$m and 16 to 22 $\mu$m 
as a function of redshift for the SED templates described 
above, where fluxes are given per unit frequency (i.e. flux density).

  \begin{figure}
\resizebox{7cm}{!}{\includegraphics{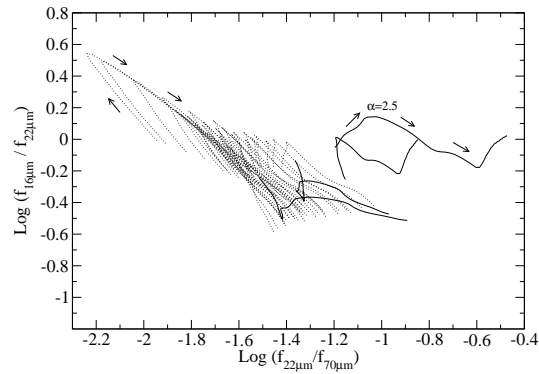}}
 \caption{The separation of silicate-break galaxies and 
low-$z$ quiescent star-forming galaxies. Dotted lines indicate 
the colours of ULIRGs at $z=1$ -- 2. Solid lines are for 
quiescent star-forming galaxies at $z<1$. Arrows indicate the 
direction of the change in colours with increasing redshift. 
The model of Dale et al. with $\alpha = 2.5$ (coolest one) is 
indicated explicitly. 
}
 \label{colcol1}
\end{figure}

For the 8 to 24 $\mu$m flux ratio, the colour bump around $z=1.5$ 
due to silicate absorption is not prominent. 
The flux ratio can vary with a rather large scatter at different redshifts, 
which makes the silicate-break selection difficult. 
This is because the stellar light mainly contributes to the 8 $\mu$m 
flux, while dust emission contributes at 24 $\mu$m. 
Also, note that quiescent star-forming galaxies at 
$z< 0.5$ have a similar flux ratio to that of ULIRGs at $z\sim 1.5$. 
Obviously, this flux ratio is not useful for the selection of silicate-break 
galaxies. 

We can expect a weaker variation of the flux ratio as a function of 
redshift, when the dust emission 
contributes in both the selected bands. This is the case for the 
flux ratio of 24 to 70 $\mu$m for the {\it Spitzer} band. 
From the flux variation predicted by 
the SED model, galaxies with $\log (f_{24}/f_{70}) \la -1.7$ could be 
silicate-break galaxies at $z\sim 1.5$. 
This selection is effective for galaxies with relatively deep 
silicate absorption feature. 
This means that 24-to-70 $\mu$m selection is likely to be biased 
towards high optical depth. 
The contamination from quiescent star-forming galaxies is negligible. 

With the IRS peak-up imagers at 16 and 22 $\mu$m, we could make 
more complete sample of silicate-break galaxies than that of 
24-to-70 $\mu$m selection, i.e. free from the bias in optical depth. 
If we adopt the criterion of $\log (f_{16}/f_{22}) \ga -0.1$, 
most of the SED templates of ULIRGs satisfy this criterion at $z=1$ -- 2. 
Although this selection suffers from some contamination from 
quiescent star-forming galaxies with cold dust temperatures at $z<1$, 
 it can be removed by using additional photometry 
at 70 $\mu$m. In Figure \ref{colcol1}, we show the colour-colour 
diagram with the two IRS imagers and 70 $\mu$m. 
The 22-to-70 $\mu$m ratio is higher for more quiescent galaxies, 
which is distinguishable from that of ULIRGs at $z=1$ -- 2. 
The selection with the 16-22 $\mu$m ratio also includes ULIRGs 
at $z>3$, which are interesting objects as well. The fraction 
of such high-$z$ ULIRGs would be small with the current sensitivity limits.





\subsection{Case for {\it ASTRO-F}}
{\it ASTRO-F} has a more comprehensive set of photometric bands at MIR 
wavelengths, compared to {\it Spitzer}. The Infra Red Camera (IRC) 
instrument on {\it ASTRO-F} has 3 channels each comprising of 3 photometric 
bands (Wada et al. ~\shortcite{wada03}). 
The IRC-NIR has bands at 2.4, 3.2, 4.3$\mu$m. The IRC-MIR-S has bands at 7, 9, 11$\mu$m. The IRC-MIR-L  has bands at 15, 20, 24$\mu$m. The bands at 9 and 20$\mu$m are considerably wider than the others. Thus among the three channels of 
the IRC, the MIR-L channel has the most useful bands to cover the silicate absorption in 
galaxies at $z>1$; i.e. 15 $\mu$m (L15), wide 20 $\mu$m (L20W), and 24 $\mu$m (L24).

\begin{figure}
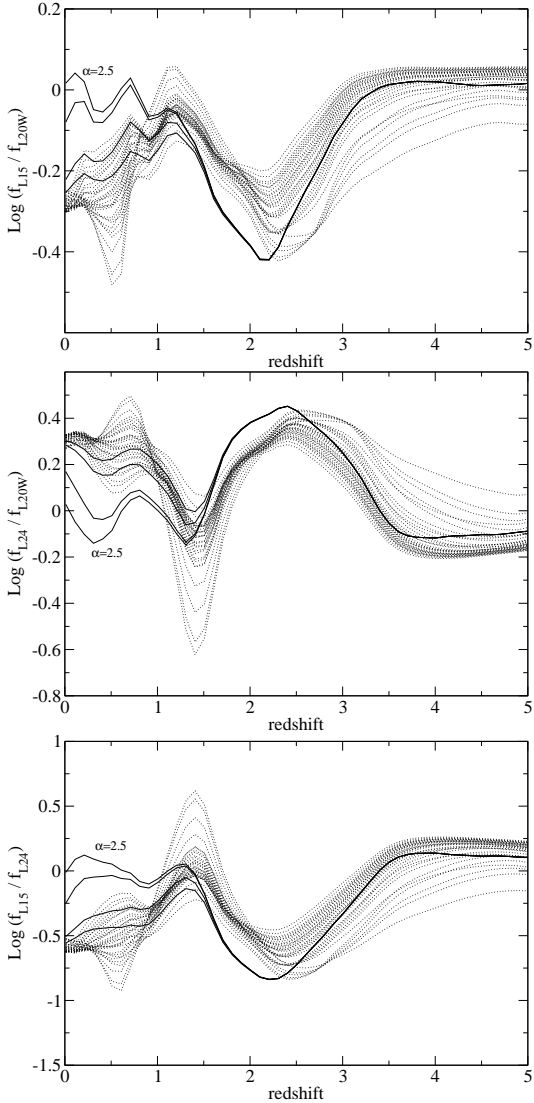

\resizebox{7cm}{!}{\includegraphics{figure/astrof_L15_L20W_smc.eps}}
\resizebox{7cm}{!}{\includegraphics{figure/astrof_L24_L20W_smc.eps}}
\resizebox{7cm}{!}{\includegraphics{figure/astrof_L15_L24_smc.eps}}
 \caption{The flux ratios with {\it ASTRO-F} bands as a function 
of redshift. The same line style is used as in Figure \ref{spitzer}.}
 \label{astrof}
\end{figure}

\begin{figure}
\resizebox{7cm}{!}{\includegraphics{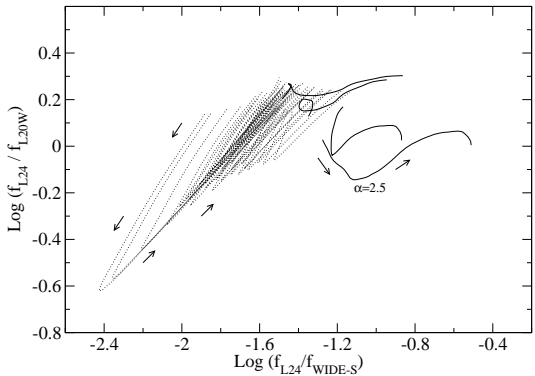}}
 \caption{
The same as Figure \ref{colcol1}, but for {\it ASTRO-F} bands. 
The WIDE-S band is the wideband filter at 80 $\mu$m. 
}
 \label{colcol2}
\end{figure}

In Figure \ref{astrof}, we show three flux ratios as a function of 
redshift. The flux ratio of L24 to L20W has the most prominent colour 
bump at $z\sim$1.5 due to the silicate absorption. Galaxies with 
$\log (f_{\mathrm{L24}}/f_{\mathrm{L20W}}) \la 0$ could be 
silicate-break galaxies. In this colour cut, most of the galaxies showing the
silicate absorption feature could be selected, 
irrespective of the optical depth. 
As in the case of the selection with the IRS peak-up imagers, there 
would be some contamination from 
quiescent star-forming galaxies with cold dust temperatures at $z<1$. 
Again, this contamination can be removed with additional photometry 
at FIR wavelengths as shown in Figure \ref{colcol2}.
Also note that this selection will include a small fraction of ULIRGs at $z>3$.

\subsection{Test with observed spectra} \label{sec:obssed}
Takagi et al. (2003b) perform a detailed comparison of the SED 
model with the observations of Arp220 and M82. They find that 
the depth of the silicate absorption features in Arp220 and M82, 
along with the UV-submm SED, 
are reproduced by the SMC and LMC dust models, respectively. 
The detailed observed spectra of silicate absorption 
features are slightly different from that of the model, which 
can be attributed to the uncertainty of the optical properties of 
dust grains. Also note that some ULIRGs show strong absorption 
features of molecules, such as CO gas and water ice (Spoon et al. 2004), 
which are not taken into account in the models. 
Thus, we need a test of derived selection criteria by using 
observed spectra. 

Here we test our selection criteria by using 8 observed 
spectra obtained with {\it ISO} and {\it Spitzer}, i.e. 
Arp220, M82 (Elbaz et al. 2002), NGC1068 (Sturm et al. 2000), 
IRAS F00183-7111 (Spoon et al. 2004a), NGC4418 (Spoon et al. 2004b), 
UGC5101, Mrk1014, and Mrk463 (Armus et al. 2004)\footnote{
Since the digital form of the spectra in Armus et al. (2004) is 
not available to the public, we used the data trace software 
`DataThief II' for the electric version of the published papar. 
The error on this process is negligible for our study.}. 
Among these galaxies, AGN-continuum dominates the MIR emission 
in NGC1068 (typical Seyfert 2), Mrk1014 (infrared luminous QSO) 
and Mrk463 (Seyfert 2). Both NGC4418 and IRAS F00813-7111 have 
strong silicate absorption features, 
although the strong radio emission indicate the presence of AGN. 
By nature, this sample is not complete in any statistical sense. 

\begin{figure}
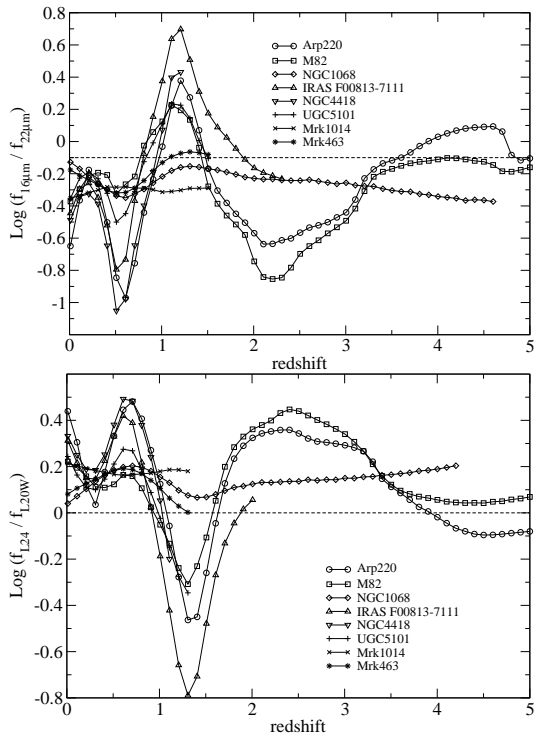

\resizebox{7cm}{!}{\includegraphics{figure/spitzer_IRS_obs.eps}}
\resizebox{7cm}{!}{\includegraphics{figure/astrof_L24_L20W_obs.eps}}
 \caption{
The flux ratios for {\it Spitzer} band and {\it ASTRO-F} 
band expected from the observed spectra of nearby galaxies. 
Dashed holizontal lines indicate the selection criteria of silicate-break 
galaxies. 
}
 \label{obs}
\end{figure}

In Figure \ref{obs}, we show the 16-to-22 $\mu$m flux ratio for 
{\it Spitzer} and the L24-to-L20W for {\it ASTRO-F} as a function 
of redshift based on the observed spectra. 
The selection criteria based on the model analysis result in the 
selection of 5 out of 8 galaxies as silicate-break galaxies. 
As expected, these selections miss the continuum-dominated galaxies, 
which have almost constant flux ratio and cause no contamination. 

Note that the variation of flux ratios are similar to 
those of the SED models. This suggests that these flux ratios are 
insensitive to the detailed features in the MIR spectra, such as 
molecular absorptions. Also, the selection criteria 
would be less sensitive to any uncertainty in 
the dust model, such as the detailed shape of silicate feature. 

\begin{figure}
\resizebox{7cm}{!}{\includegraphics{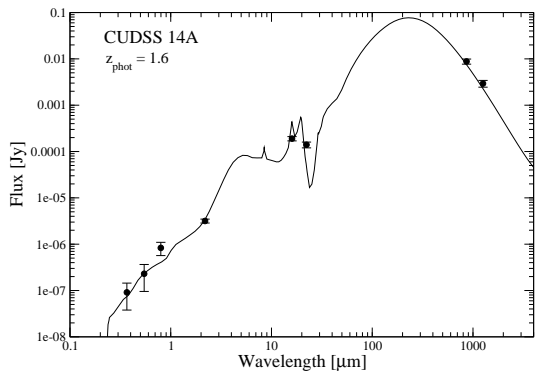}}
 \caption{
The result of SED fitting for CUDSS 14A. The fluxes at 
optical-NIR bands, the IRS bands, 850 $\mu$m and 1.3 mm 
are taken from Lilly et al. (1999), Charmandaris et al. (2004), 
Eales et al. (1999), and Gear et al. (2000), respectively. 
}
 \label{14a}
\end{figure}

In Charmandaris et al. (2004), there are two galaxies which clearly 
satisfy the silicate-break selection, i.e. object 9 and 11 in their table. 
The flux ratio of 16 to 850 $\mu$m suggests that the object 11 
(CUDSS 14A) is likely to be at $z>1$. We perform the SED fitting 
for this galaxy by using the same SED models as those in Takagi 
et al. (2004). The best-fitting SED model suggests that this galaxy 
is a silicate-break galaxy at $z=1.6$ as shown in Figure \ref{14a}.
We suggest that this galaxy is the first promising candidate of 
a silicate-break galaxy.

\section{Expected number of silicate-break galaxies}\label{sec:numbers}

\begin{figure}
  \resizebox{\hsize}{!}{\includegraphics{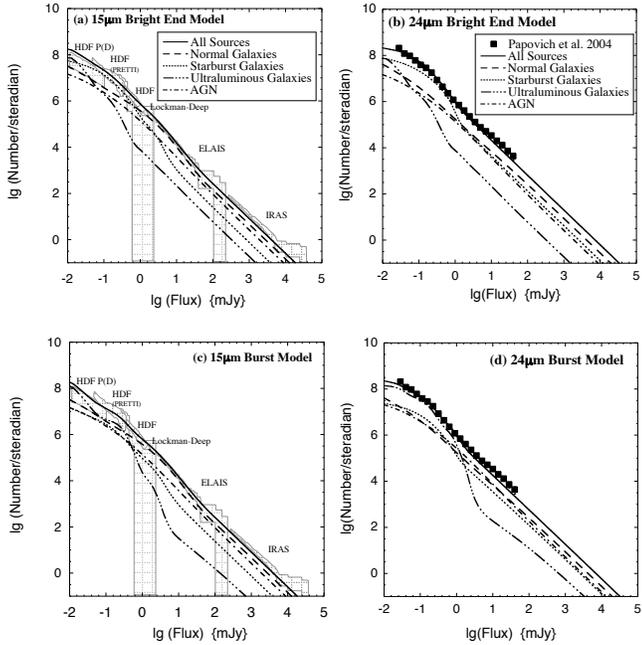}}
 \caption{Model fits (sources per steradian)  to the ISO 15$\mu$m counts and the {\it Spitzer}  24$\mu$m for the {\it Bright End Model} and the  {\it Burst Model} described in the text. ISO 15$\mu$m counts are from Hubble Deep Field (P(D) analysis - Oliver et al. (1997), HDF counts derived using the PRETI method - Aussel et al. (1999), Lockman Hole - Elbaz et al. (1998), ELAIS - Serjeant et al. (2000) and shifted IRAS counts. 24$\mu$m counts are from Papovich et al. (2004)
}
 \label{observed}
\end{figure}

We use the galaxy evolution models of Pearson ~\shortcite{cpp04b} to 
predict the numbers of silicate-break galaxies expected in the 
mid-infrared surveys that are or will be conducted by {\it Spitzer}  
and {\it ASTRO-F}. These models use the type dependent luminosity 
functions derived from 
the ISO-ELAIS survey 
(Rowan-Robinson et al. ~\shortcite{mrr04}) 
of Pozzi et al. ~\shortcite{pozzi04} and Matute et al.~\shortcite{matute02} 
to represent the normal and starburst/ULIRG and AGN populations 
respectively. We investigate two particular models,  in order to 
predict the numbers of silicate-break galaxies. The models are 
categorized by their dominant populations of 
starburst (M82 like)  and ULIRG (Arp 220 like) sources respectively.  
The first (starburst dominated) model is referred to as the 
{\it Bright End Model}, and broadly follows the evolutionary 
scenario of Pearson \& Rowan-Robinson ~\shortcite{cpp96}. This 
model assumes power law evolution in both luminosity and density of 
the forms $(1+z)^{k}$ for the starburst, AGN and ULIRG populations 
respectively. The second (ULIRG dominated) model is referred to as 
the {\it Burst Model}, and broadly follows the evolutionary scenario 
of Pearson  ~\shortcite{cpp01}. This model assumes a similar power law 
evolution in luminosity, for the starburst and AGN populations and an 
initial exponential burst + power law evolutionary scenario  for the 
ULIRG population. In both models the normal galaxies are assumed to be 
non-evolving. See Pearson ~\shortcite{cpp04b} for details of the models. 
These models provide good fits to both the 15$\mu$m ISO counts and
 the {\it Spitzer}  counts at 24$\mu$m  
(Papovich et al.  ~\shortcite{papovich04}, 
Oliver et al. ~\shortcite{oliver97},  
Aussel et al. ~\shortcite{aussel99}, 
Elbaz et al.  ~\shortcite{elbaz98}, 
Serjeant et al.  ~\shortcite{serjeant00}). 
These model fits are shown in Figure \ref{observed}. 

For the purpose of this study, to model the ULIRG population (i.e. the potential silicate-break galaxies) 
we have used two SED templates from the library of Takagi (2003a) rather than that of Arp220 which has a rather anomalous farinfrared-midinfrared ratio. 
These SEDs are modelled on the local IRAS galaxies, IRAS 15250 and 
IRAS 12112, for which the model gives an excellent fit to 
the observed multi-band data, specifically at NIR-submm wavelengths. 
The SED of IRAS 15250 and IRAS 12112 are 
assumed to be representative of infrared luminosities of  
($L_{IR} \ga 10^{11} L\sun $) and ($L_{IR} \ga 10^{12} L\sun $), respectively. 

\begin{figure}
  \resizebox{\hsize}{!}{\includegraphics{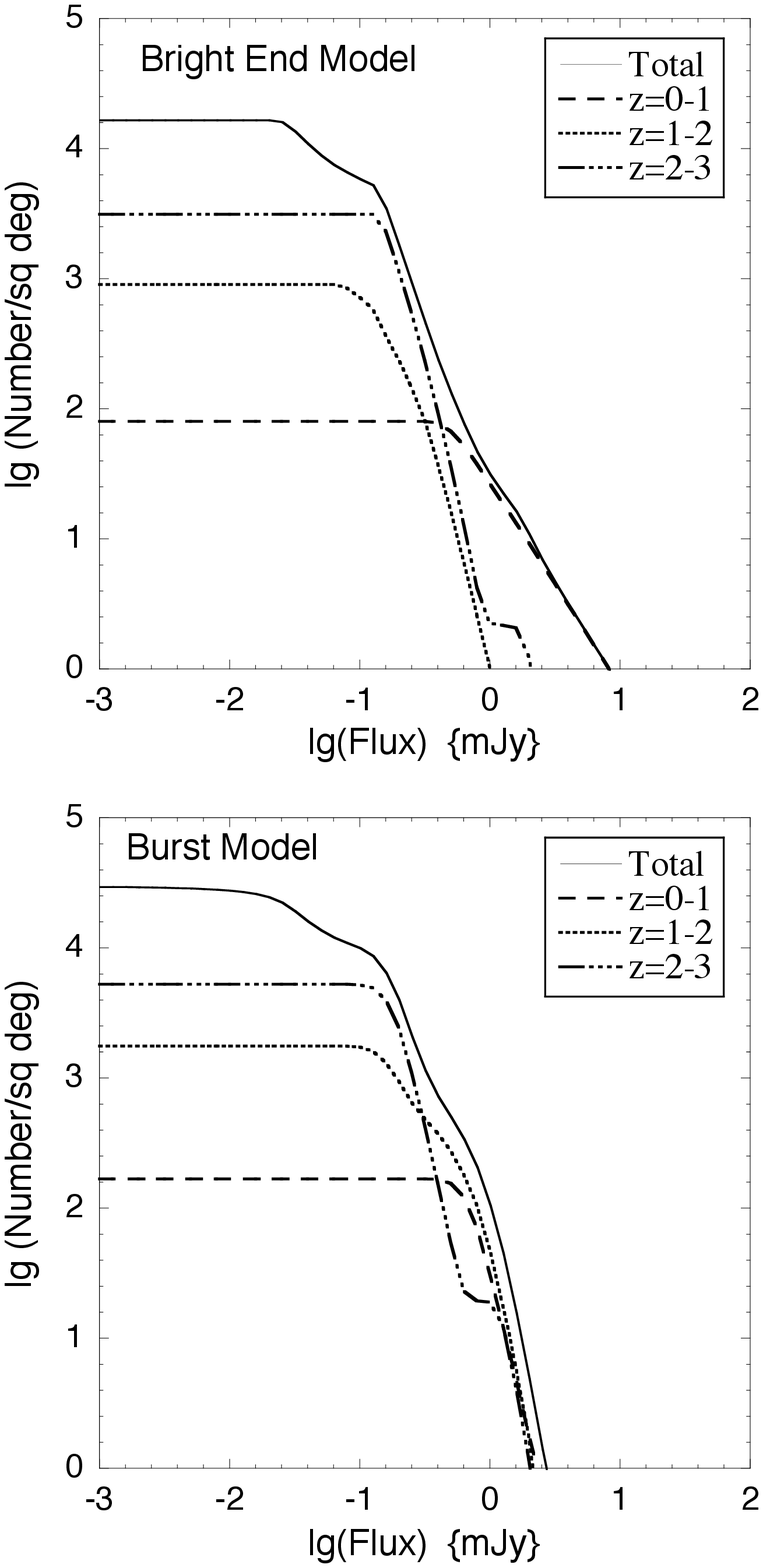}}
 \caption{
Predicted number of ULIRGs for the Bright End and Burst
evolutionary models, as a function of redshift (sources per square 
degree).
Numbers are divided into redshift bins of $z=0$ -- 1, 1 -- 2, and 
2 -- 3. The selection criteria for the silicate-break effect dictates 
that the interesting sources will be ULIRGs between $z = 1$ -- 2. 
}
 \label{counts}
\end{figure}

The selection criteria for silicate-break galaxies dictates 
that the interesting sources will be ULIRGs between $z=1$ -- 2. 
Hence in Figure \ref{counts} we plot the number of sources per 
square degree as a function of flux at 24$\mu$m in 3 redshift 
bins of $z=0$ -- 1, 1 -- 2 and 2 -- 3.
For the {\it Spitzer}  24$\mu$m band, the number of potential 
silicate-break galaxies are predicted to be $\sim900$ and $\sim1500$/sq.deg 
for the {\it Bright End} model and the {\it Burst} model 
respectively. All the silicate-break galaxies selected at 24$\mu$m 
should be detected at the confusion limit of the 70$\mu$m surveys 
with {\it Spitzer}  
(Lonsdale et al. ~\shortcite{lonsdale03}, ~\shortcite{lonsdale04}). 

{\it ASTRO-F} will have a more comprehensive set of bands in the 
mid-infrared, having 3 bands sensitive to the silicate-break galaxies 
(15, 20, 24$\mu$m) with its' IRC instrument. Present survey strategies 
for the IRC include both a deep survey and a shallow survey around 
the north ecliptic pole (NEP) region with the sensitivity of 
$\sim 50$ and $\sim 170$ $\mu$Jy ($5\sigma$) 
at L20W band, respectively (Pearson et al. ~\shortcite{cpp01b}. 
~\shortcite{cpp04c}). 
The survey area of the deep and the shallow survey would be 
$\sim$0.5 and $\ga 3$ deg$^2$, respectively 
(Pearson et al. ~\shortcite{cpp01b}, 
Matsuhara \& Pearson~\shortcite{matsuhara04}, 
Pearson \& Matsuhara~\shortcite{cpp04c}). 
%
For the deep NEP survey similar numbers to 
the  {\it Spitzer}  24$\mu$m results for the 2 evolutionary 
models are expected in all 3 bands of the MIR-L since we are effectively 
detecting all ULIRGs between $z=1$ -- 2 at these depths. For the 
shallow pointing survey, we predict $\sim400$ and $\sim1500$ 
candidates per square degree at 20 $\mu$m for
the {\it Bright End} model and the {\it Burst} model, respectively. 

The actual number of silicate-break galaxies would be somewhat 
smaller than the predicted number of ULIRGs at $z=1$ -- 2, 
since the possible contribution from the AGN continuum make the silicate 
absorption less prominent. 
Note that the fraction of AGN-dominated galaxies, i.e Seyfert 1 and 2,
in nearby ULIRGs are $\sim 30$~\% (Sanders \& Mirabel 1996). 
X-ray follow-up observations of deep {\it ISO} surveys in the 
HDF-N and the Lockman Hole consistently 
suggest that $\sim 25$~\% of {\it ISO}-detected 
MIR sources are likely to be AGN-dominated (Fadda et al. 2002).  
X-ray properties of the silicate-break population and other ULIRGs 
would be very important to study the obscuration of AGNs by dust 
at high redshifts. 

The observed number of silicate-break galaxies will place a 
strong lower limit on the number of ULIRGs at this redshift range. 
From the above 
predictions and our silicate-break selection criteria, 
it will be possible to discriminate between the models above and 
more extreme evolutionary models (e.g. Dusty E/S0 
models ~\cite{xu03}, cirrus dominated models ~\cite{mrr01}). 
From Figure~ \ref{counts}  it can be seen that even 
the  {\it Bright End}  and  {\it Burst} evolutionary 
models differ by factors of 2 -- 3 in their predictions for 
ULIRGs between $z\sim 1$ and 2.

\section{Discussion and conclusions}\label{sec:conclusions}

We have investigated the photometric pre-selection method of 
high-$z$ ULIRGs at $z= 1$ -- 2, 
by focusing on the silicate absorption feature at 
$9.7 \mu$m in the rest-frame. 
This technique requires observations only at MIR 
wavelengths, and is therefore very efficient and self-consistent. 
Applying our silicate-break technique to the sample of Charmandaris et al. (2004) we have tentatively identified the first candidate silicate-break galaxy at  $z=1.6$ .

With {\it Spitzer}, we could select ULIRGs at $z\sim 1.5$ as 
silicate-break galaxies 
with $\log (f_{24}/f_{70}) < -1.7$. This colour cut is applicable 
to the selection of optically thick ULIRGs; i.e. the selection 
is biased towards high optical depth, although the sample is free 
from the contamination from quiescent star-forming galaxies. 
A more complete selection of 
silicate-break galaxies is possible with the IRS peak-up imagers 
with additional photometry at 70 $\mu$m to remove the contamination 
from quiescent star-forming galaxies at $z<1$. 

With {\it ASTRO-F}, a similar 
selection criterea of silicate-break galaxies to that of the IRS imagers 
is possible with L24 and L20W bands. The advantage of {\it ASTRO-F} 
is the larger field-of-view (100 arcmin$^2$) for these bands, compared 
to that of the IRS peak-up imagers, allowing for larger samples of 
silicate-break galaxies. 

The proposed selection criteria is useful to construct samples 
of heavily obscured galaxies at $z= 1$ -- 2. The majority of such galaxies 
are likely to be powered by starbursts rather than AGN. 
Therefore, such samples could be used to place strong limits on the 
star formation activity at $z=1$ -- 2. 
Note that spectroscopic redshifts might be difficult to measure for such 
heavily obscured galaxies, which are expected to be faint at optical 
wavelengths. Furthermore, the expected redshift range 
of silicate-break galaxies overlaps with the so-called 
`{\it redshift desert}', in which strong emission lines are not accessible 
from the ground. Thus, the proposed silicate-break selection method could play 
important role to investigate the nature of infrared galaxies 
at $z>1$ bridging the gap between the IRAS-ISO and SCUBA populations. 

In the {\it ASTRO-F} NEP surveys, 
we expect to detect $\sim400$ -- 1500 candidates of silicate-break 
galaxies depending on the evolutionary model. This large sample 
will allow us to derive important 
characteristics of ULIRGs at $z=1$ -- 2, such as 
their luminosity and spatial correlation function. Such data sets 
can also discriminate or tightly constrain multi-component number count models.

The available filters on {\it Spitzer} and 
{\it ASTRO-F} are found to be too broad to use individual 
PAH emission features for photometric pre-selections. 
Similarly, we find that the {\it ISO} 15 $\mu$m band is too broad 
to produce the colour anomaly for galaxies at $z\sim 0.5$, 
owing to the silicate absorption. 
Thus, using the silicate-break method, {\it Spitzer} and 
{\it ASTRO-F} can provide the first opportunity to 
pre-select high-$z$ infrared galaxies with MIR bands. 

\section*{Acknowledgments}
We thank S. Serjeant for useful and stimulating discussions. Also, we thank
H. Hanami for useful comments, which improved the paper. We would like to
thank D. Elbaz and H.W.W. Spoon for providing us with the 
observed spectra of galaxies. 
We are also grateful to D. Dale for distributing the SED
template on the Web. We thank the referee for stimulating comments. 
CPP acknowledges a Fellowship to Japan from the
European Union.


\end{document}